\documentclass[11pt]{article}
\usepackage{geometry}                
\usepackage{graphicx}
\usepackage{amssymb}
\usepackage{mathtools}
\usepackage{epstopdf}
\usepackage{verbatim}
\usepackage{color}
\usepackage{slashed}
\usepackage{bbold}
\usepackage{fullpage}
\usepackage{authblk}
\usepackage{float}
\usepackage[title]{appendix}
\usepackage{caption}
\usepackage{subcaption}
\usepackage[colorlinks, citecolor=red, linkcolor=blue]{hyperref}

\restylefloat{table}
\restylefloat{figure}

\newcommand{\beq}{\begin{equation}}
\newcommand{\eeq}{\end{equation}}
\newcommand{\bmat}{\begin{pmatrix}}
\newcommand{\emat}{\end{pmatrix}}
\newcommand{\bal}{\begin{align}}
\newcommand{\eal}{\end{align}}

\newcommand{\Order}{\mathcal{O}}
\newcommand{\sg}{\tilde{g}}
\newcommand{\sq}{\tilde{q}}

\DeclareMathOperator{\TeV}{TeV}
\DeclareMathOperator{\GeV}{GeV}

\begin{document}

\title{\begin{flushright}\small{MCTP-14-26}\end{flushright} 
~\\
~~\\
\huge\bf{Superpartners at LHC and Future Colliders:}\\ \LARGE{Predictions from Constrained Compactified M-Theory}}

\author{\normalsize\bf{Sebastian~A.~R.~Ellis\footnote{sarellis@umich.edu}~}}
\author{\normalsize\bf{Gordon~L.~Kane\footnote{gkane@umich.edu}~}}
\author{\normalsize\bf{Bob Zheng\footnote{byzheng@umich.edu}}}
\affil{\it{Michigan Center for Theoretical Physics (MCTP),}\\ \it{Department of Physics, University of Michigan}, \\ \it{Ann Arbor, MI 48109, USA}}
\date{\small{\today}}
\maketitle
\begin{abstract}
We study a realistic top-down M-theory compactification with low-scale effective Supersymmetry, consistent with phenomenological constraints. A combination of top-down and generic phenomenological constraints fix the spectrum. The gluino mass is predicted to be about 1.5 TeV. Three and only three superpartner channels, $\sg \sg$, $\chi_2^0 \chi_1^\pm$ and $\chi_1^+ \chi_1^-$ (where $\chi_2^0, \chi_1^\pm$ are Wino-like), are expected to be observable at LHC-14.  We also investigate the prospects of finding heavy squarks and Higgsinos at future colliders. Gluino-stop-top, gluino-sbottom-bottom associated production and first generation squark associated production should be observable at a 100 TeV collider, along with direct production of heavy Higgsinos. Within this framework the discovery of a single sparticle is sufficient to determine uniquely the SUSY spectrum, yielding a number of concrete testable predictions for LHC-14 and future colliders, and determination of $M_{3/2}$ and thereby other fundamental quantities.
\end{abstract}

\newpage

\section{Introduction}
\label{Intro.SEC}

If Supersymmetry (SUSY) is considered as an effective theory of a well-motivated ultraviolet (UV) completion, concrete predictions can be made about the Supersymmetric particle (sparticle) spectrum that is to be expected at the TeV scale. In a top-down approach, the Supersymmetry breaking parameters that in a bottom-up model would be free are instead set by high-scale dynamics. This leads to concrete, testable predictions for sparticle masses, as opposed to treating sparticle masses as free parameters in a multidimensional parameter space.

In this note, we illustrate the power of this top-down approach for a particularly well-motivated UV completion of Supersymmetry: M-theory compactifications on manifolds with $G_{2}$ holonomy \cite{Papadopoulos:1995da,Acharya:1998pm,Acharya:2000gb,Atiyah:2001qf,Acharya:2001gy}. We call this framework the $G_{2}$-MSSM, following earlier papers which have studied its phenomenological properties \cite{Acharya:2006ia, Acharya:2007rc, Acharya:2008zi, Acharya:2008hi}. Due to the rigidity of top-down theoretical constraints from moduli stabilization and Supersymmetry breaking, the sparticle spectrum is completely fixed by electroweak symmetry breaking (EWSB), once both the gravitino mass $M_{3/2}$ and the superpotential $\mu$-term are specified. Furthermore, the measured value of the Higgs mass picks out a particular slice in the $(\mu, M_{3/2})$ plane; thus \emph{all} sparticle masses can be fixed by determining either $\mu$ or $M_{3/2}$. 

Given minimal assumptions regarding the geometry of the compact $G_2$-manifold, $M_{3/2}$ is approximately calculable once the moduli are stabilized and SUSY is broken. This results in a central value of $M_{3/2} = 35$ TeV. Imposing EWSB and Higgs mass constraints, $M_{3/2} = 35$ TeV corresponds to a gluino mass of $1.5$ TeV. The result $M_{3/2} = 35$ TeV can be altered if UV threshold corrections to the non-perturbative superpotential are non-negligible. Allowing a wider range of $20 \, \mathrm{TeV} < M_{3/2} < 50\, \mathrm{TeV}$ due to currently unknown UV corrections gives a range of gluino masses, $1 \, \mathrm{TeV} < M_{gluino} < 2$ TeV. Thus a gluino mass within this range can be taken as a prediction of the compactified M-theory framework.

For concreteness we will focus on the central value $M_{3/2} = 35$ TeV, which gives a benchmark spectrum with a $1.5$ TeV gluino. All sfermion and heavy Higgs masses are of $\mathcal{O}(M_{3/2})$, while the Wino(Bino)-like lighter gauginos have masses of $615 (450)$ GeV. The hierarchy between gauginos and $M_{3/2}$ follows from the dynamics of moduli stabilization. Specifically, both the hidden sector meson and moduli F-terms contribute to the gravitino mass $M_{3/2}$, while only the moduli F-terms contribute to gaugino masses. The moduli F-terms are suppressed with respect to hidden sector meson F-terms by about $\alpha_{GUT} \approx 1/25$, resulting in $\lesssim$ TeV gaugino masses despite $M_{3/2}$ being tens of TeV \cite{Acharya:2008zi}. As will be discussed below, this benchmark spectrum is not constrained by LHC-8. Both gluino pair production \emph{and} direct electroweak gaugino production should yield discoveries with $\lesssim 300$ fb $^{-1}$ of LHC-14 data, particularly since electroweak gauginos yield distinctive signatures through $\chi^0_2 \rightarrow \chi^0_1\, +  h$ which has a nearly $100\%$ branching ratio. This bencmark spectrum predicts $\mu \approx 1.4$ TeV, leading to heavy Higgsinos which are out of reach of LHC-14.

We will also discuss implications of the $G_2$-MSSM for future colliders. Taking the benchmark spectrum, we will show that $1.5$ TeV Higgsinos are accessible at both 50 and 100 TeV colliders. Furthermore, the heavier squarks are also accessible at 100 TeV colliders, with hundreds of squark-gluino associated production events expected with $\gtrsim 1000$ fb$^{-1}$ of data. In particular, we point out for the first time (as far as we know) that for a gluino mass of $\sim$ TeV, the cross-section for associated stop-gluino-top production $p p \rightarrow \tilde{g} + \tilde{t}_1+ t$, and potentially even sbottom-gluino-bottom $p p \rightarrow \tilde{g} + \tilde{b}_1 + b$, can be sizeable at 100 TeV colliders for stops and sbottoms lighter than $\sim 20$ TeV. This is especially relevant for MSUGRA-like theories with universal scalar masses at the GUT scale, in which third generation squarks are expected to be lighter due to RGE effects.

Thus the benchmark spectrum has the remarkable feature that Higgsinos and squarks are accessible at future colliders, despite the scale of SUSY breaking $M_{3/2}$ being in the tens of TeV range. Furthermore, the constrained relationship between SUSY breaking parameters implies that within the $G_2$-MSSM, \emph{the discovery of a single sparticle is enough to actually measure $M_{3/2}$}. Given the discovery of a single sparticle, the rest of the $G_2$-MSSM spectrum is determined uniquely, resulting in a multitude of additional predictions which can be readily confirmed or falsified in ongoing and upcoming collider experiments. This point highlights the potential power of top-down approaches in greatly reducing the naive parameter space of Supersymmetry. We will provide details concerning these points in later work; the results can basically be seen from equations (\ref{GUTparams}, \ref{gravitinoeq}) and Figure \ref{Fig1} in Section \ref{Features.SEC} below.

The paper is organised as follows. Section \ref{Features.SEC} gives a brief overview of the $G_2$-MSSM framework, and discusses how the benchmark spectrum corresponding to $M_{3/2} = 35$ TeV is obtained. Readers interested only in experimental predictions should skip this section. Section \ref{Spect.SEC} provides the sparticle spectrum and relevant branching ratios corresponding to the benchmark spectrum obtained in Section \ref{Features.SEC}. Predictions for LHC-14 are given in Section \ref{LHC.SEC}, while predictions for future colliders are given in Section \ref{FCC.SEC}. Finally, we summarise the results in Section \ref{Conclusion.SEC}. 

\section{Theoretical Framework}
\label{Features.SEC}

We begin by reviewing some features and successes of the compactified M-theory framework. $G_2$-compactifications of M-theory provide a natural setting for full moduli stabilisation and broken $\mathcal{N}=1$ Supersymmetry in a deSitter vacuum, while also solving the gauge hierarchy problem \cite{Acharya:2006ia, Acharya:2007rc, Acharya:2008zi, Acharya:2008hi}. Additionally, EDM \cite{Kane:2009kv, Ellis:2014tea} and flavour constraints (such as $B_s \to \mu \mu$) \cite{Kadota:2011cr} are avoided. 
However, it may not explain $(g-2)_\mu$, while the strong CP problem is solved by the axionic components of moduli fields \cite{Acharya:2010zx}. The presence of late-decaying moduli results in a non-thermal cosmological history which solves the moduli problem and the gravitino problem \cite{Acharya:2008bk}. Since the baryon asymmetry and the dark matter both arise from moduli decay (including axions), the ratio of baryonic matter to dark matter is calculable \cite{Kane:2011ih}. R-parity conservation is expected \cite{Acharya:2014vha}; thus consistency with the observed DM relic abundance implies that the visible sector LSP will decay to hidden sector DM particles \cite{DMPaper}\footnote{For non-thermal cosmologies, a stable Bino-LSP will overclose the universe, while a stable Wino-LSP is in tension with indirect detection contraints \cite{Fan:2013faa,Cohen:2013ama}.}. $\mu$ is incorporated into the theory following the proposal of Witten \cite{Witten:2001bf} and including effects of moduli stabilisation \cite{Acharya:2011te}. Calculations in this framework anticipated the mass and decay branching ratios of the Higgs boson observed at the LHC \cite{Kane:2011kj}.

A key result of the aforementioned references is that once moduli stabilization breaks Supersymmetry and cancels the vacuum energy at tree level, the relationship between SUSY breaking parameters becomes very constrained. Upon moduli stabilization and SUSY breaking, the soft breaking parameters at the renormalization scale $Q \sim M_{GUT}$ are given by \cite{Acharya:2008hi}:\begin{align}\notag  & m_0^2 \approx M_{3/2}^2\left(1-C\right), \, \,\, A_0^2 \approx 1.5 M_{3/2} \left(1- C \right),\\ \label{GUTparams} &  M_a \approx \left[-0.032 \eta  +\alpha_{GUT}\left(0.034 \left(3 C_a - C^\prime_a \right) + 0.079 C^\prime_a(1 - C)\right) \right]\times M_{3/2}\end{align} where $C_a = (0, 2, 3)$ and $C_a^\prime = (33/5, 7, 6)$. $m_0$ and $A_0$ are universal soft scalar masses and trilinears, and ``$C$" parameterizes higher order K\"ahler potential corrections arising from higher dimensional operators as defined in Appendix \ref{moduli}. The quantity $\eta$ parameterizes KK-threshold corrections to the unified gauge coupling; we will argue in Section \ref{benchmark} that $\eta \sim 1$, unless the geometry of the $G_2$ manifold becomes incredibly complicated. A review of moduli stabilization and SUSY breaking along with the derivation of (\ref{GUTparams}) is presented in Appendix \ref{moduli}. The hierarchy between gaugino masses and $M_{3/2}$ arises because $M_{3/2}$ feels contributions from both hidden sector meson and moduli F-terms, while $M_a$ feels contrbutions from only the moduli F-terms which are suppressed by approximately $\alpha_{GUT} \approx 1/25$ with respect to the meson F-terms. A more detailed discussion of this point is deferred to Appendix \ref{moduli}.

Thus the \emph{entire} sparticle spectrum is essentially fixed once $M_{3/2}$, $\mu$ and $C$ are specified. We will show in Section \ref{EWSBandMH} that imposing consistent EWSB along with the measured value of the Higgs mass $M_h = 125.2 \pm 0.4$ GeV \cite{Aad:2014aba,CMS:2014ega} reduces the 3D space of allowed $(M_{3/2}, \mu, C)$ values to an approximately one-dimensional space. As a result, the entire sparticle spectrum is completely determined for a given value of $M_{3/2}$. In Section \ref{benchmark} we will use top-down considerations to approximately calculate $M_{3/2}$, giving a central value of $M_{3/2} \approx 35$ TeV which we use to obtain the benchmark spectrum considered in Section \ref{Spect.SEC} and onwards.

\subsection{Imposing Constraints: EWSB and the Higgs Mass}\label{EWSBandMH}

In the previous section, we have stated that the sparticle spectrum is essentially determined by three quantities: $M_{3/2},\, \mu, \, C$, or equivalently $M_{3/2}, \mu/M_{3/2}, C$. In principle, these quantities are calculable from the full UV theory. In practice however, there are theoretical uncertainties which preclude a full top-down calculation. Instead, we will show in this section how bottom-up constraints of EWSB along with the measured Higgs mass provide two independent constraints, reducing the naive 3D region to a one-dimensional strip. This illustrates the power of combining top-down calculations with known bottom-up constraints  to increase the predictiveness of a particular theory.

A detailed discussion of how the constraints from EWSB and $M_{h} = 125.2 \, \pm 0.4 $ GeV are imposed is given in Appendix \ref{EWSB}; the result is shown in Figure \ref{Fig1}. One can see that EWSB constraints restrict the region to an approximately 2D-slice. Imposing the constraint $M_{h} = 125.2 \, \pm 0.4 $ reduces this slice to a thin band; the thickness of this band is due primarily to experimental uncertainties on $M_h$, $M_t$ and $\alpha_s$ \cite{Kane:2011kj}. The range $0 < \mu/M_{3/2} \lesssim 0.1$ is motivated by both top-down \cite{Acharya:2011te} and little hierarchy \cite{Feldman:2011ud} arguments; this is discussed in more detail in Appendix \ref{EWSB}.

\begin{figure}
\centering
\includegraphics[scale=0.6]{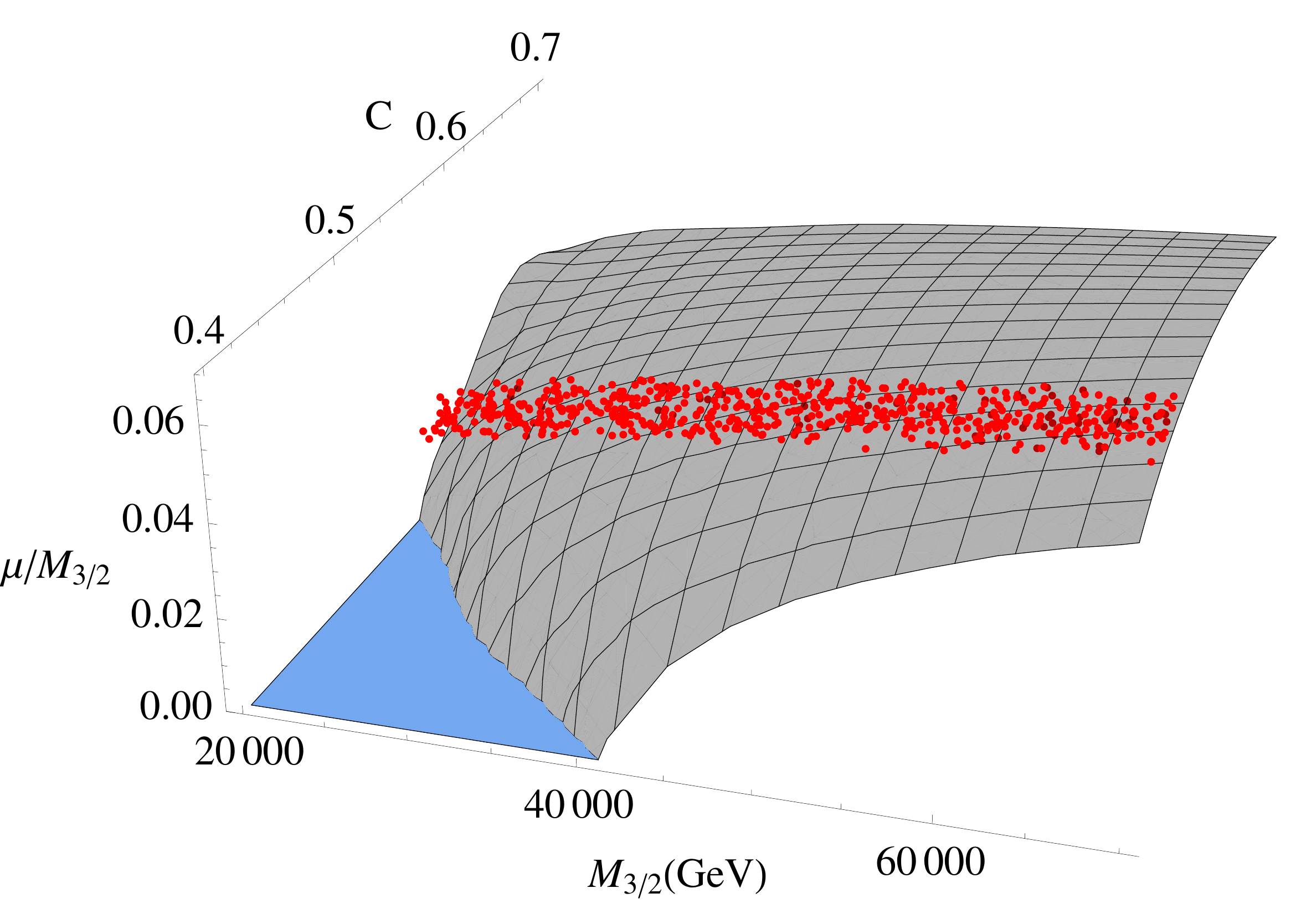}
\includegraphics[scale=0.6]{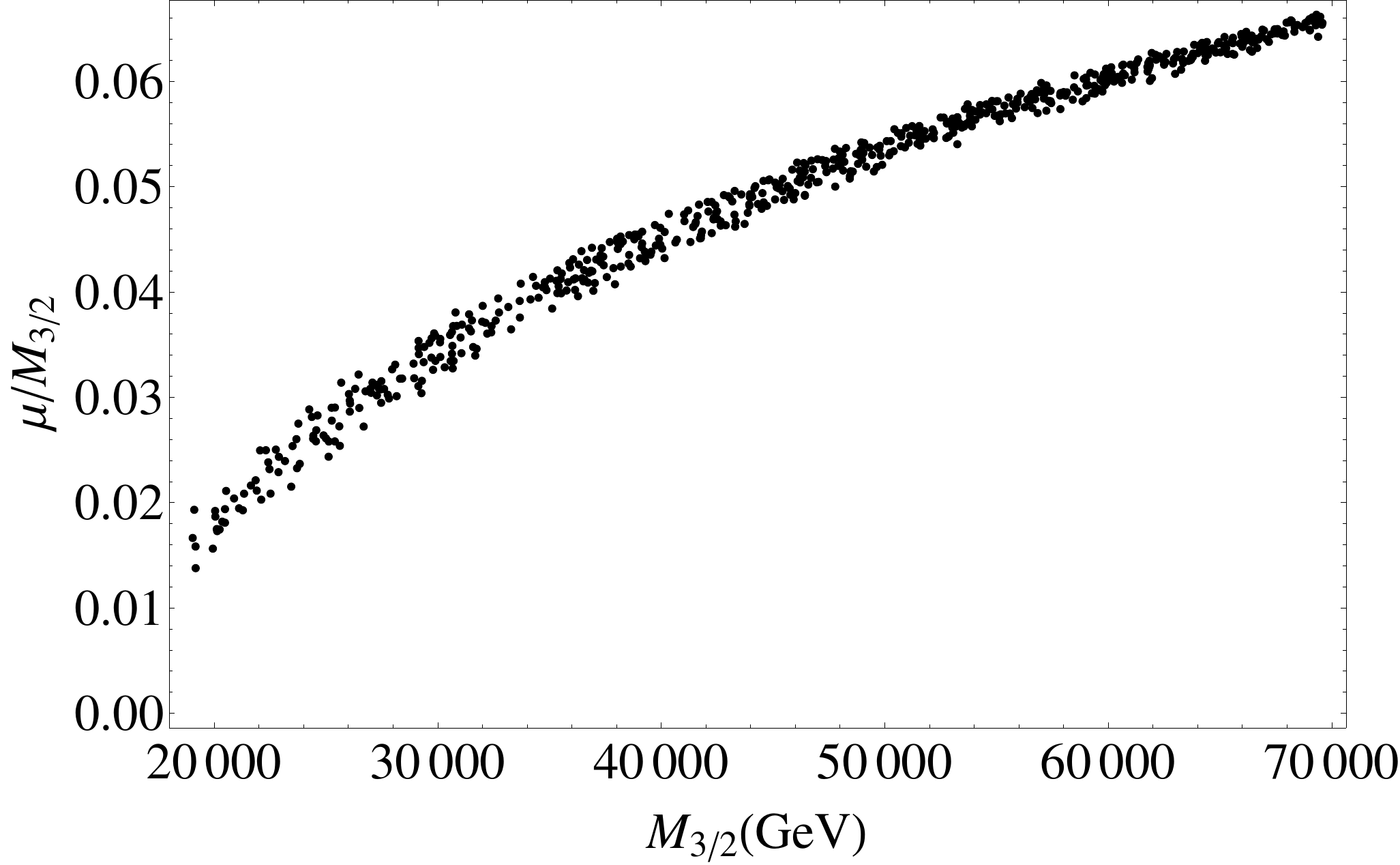}
\caption{{\bf Upper Figure}: The dark gray surface shows the slice of $M_{3/2}, c, \mu/M_{3/2}$ parameter space which satisfies EWSB, while red points also satisfy the Higgs mass constraint $M_h = 125.2 \pm 0.4$ GeV. The blue shaded region corresponds to points which are inconsistent with $\mu/M_{3/2} \lesssim 0.1$; see the Appendix for further discussion. In these plots, $\mu$ is defined at the renormalization scale $Q^2 = m_{\tilde{t}_1} m_{\tilde{t}_2}$. The mesh lines are added for perspective, and do not have any physical significance.\\
{\bf Lower Figure}: A projection of the upper figure onto the $M_{3/2}- \mu/M_{3/2}$ plane.}
\label{Fig1}
\end{figure}

Upon imposing these bottom-up constraints, the entire sparticle spectrum becomes an approximately one-dimensional strip in the original 3D space. In the next section, we take the perspective that Figure \ref{Fig1} fixes the entire spectrum in terms of $M_{3/2}$, and use equation (\ref{softparams}) to approximately compute $M_{3/2}$ to be $\approx 35 \TeV$, and obtain the associated benchmark spectrum. 

\subsection{Obtaining a Benchmark Spectrum}\label{benchmark}

Having established that the entire Supersymmetric particle spectrum is essentially determined for a given value of $M_{3/2}$, we discuss how $M_{3/2}$ can be approximately computed from the UV theory. Upon moduli stabilization and SUSY breaking, the standard Supergravity expression for the gravitino mass gives \cite{Acharya:2008hi} :
\begin{equation}\label{gravitinoeq}
M_{3/2} \approx \frac{9 \times 10^{5}}{V_X^{3/2}} \left(\frac{A_2}{Q}\right)\, \mathrm{TeV} 
\end{equation} 
where $Q$ is the rank of a hidden sector $SU(Q)$ gauge group which undergoes gaugino condensation, and $A_2$ is the corresponding non-perturbative superpotential coefficient (see Eq. (\ref{super}) in Appendix \ref{moduli}). $V_X$ is the volume of the $G_2$-manifold in 11-D Planck units. We refer interested readers to Appendix \ref{moduli} for a more detailed derivation of Eq. (\ref{gravitinoeq}).

In principle, $V_X$ is a function of the moduli fields and is thus calculable once moduli are stabilized. However the particular expression for $V_X$ in terms of moduli fields is not fully known; we can instead fix $V_X$ by ensuring that dimensional reduction to 4D gives the correct value for Newton's constant \cite{Friedmann:2002ty}. This fixes $V_X$ to be \cite{Acharya:2008hi}:\begin{equation}\label{VXeq}
V_X \approx 137.4 \, L(\mathcal{Q})^{2/3}, \hspace{4mm} L(\mathcal{Q}) = 4 q \, \sin^2 \left(\frac{5 \pi \omega}{q}\right)
\end{equation} where $L(\mathcal{Q})$ is a topological invariant which parameterizes threshold corrections from Kaluza-Klein states. This form for $L(\mathcal{Q})$ was computed in \cite{Friedmann:2002ty}, assuming that the visible sector $SU(5)$ gauge fields are compactified on a Lens space $\mathcal{Q} \cong S_3/Z_q$ in the presence of a non-trivial Wilson line. $5 \, \omega$ is an integer parameterizing the effect of the Wilson line, as will be discussed below. 

To proceed, we briefly review the motivation for considering $SU(5)$ gauge theories compactified on a Lens space $\mathcal{Q}$. In $G_2$-compactifications of M-theory, non-abelian gauge fields arise from co-dimension 4 ADE singularities and thus propogate on a 7-dimensional manifold $\mathcal{H}$ \cite{Acharya:1998pm}. In our notation, we take $\mathcal{H} \cong \mathcal{Q} \times M$ where $M$ is our Minkowski spacetime. Starting from an $SU(5)$ GUT theory, the issues of GUT breaking and doublet-triplet splitting must be resolved for a realistic model. As pointed out by Witten, both problems are elegantly solved in the presence of a non-trivial Wilson line background \cite{Witten:2001bf}, which breaks $SU(5)$ to the SM while admitting a geometric symmetry which solves doublet-triplet splitting.

The resulting symmetry is determined by the fundamental group of $\mathcal{Q}$, which for $\mathcal{Q} \cong S/Z_q$ is simply $Z_q$. The non-trivial Wilson line gives the $\int d^2 \theta \mu H_u H_d$ superpotential term charge $5 \omega$ under $Z_q$, while $\int d^2  \theta MT_u T_d$ is uncharged. However, realistic phenomenology requires non-zero $\mu$, which implies that $Z_q$  is broken once moduli with charge $5 \omega$ obtain \emph{vev}'s \cite{Acharya:2011te}. If $Z_q$ is completely broken, higher-dimensional K\"ahler potential operators will generate dangerous lepton-number violating operators which generically violate neutrino mass bounds \cite{Acharya:2014vha}. Thus $Z_q$ must be broken to a non-trivial subgroup $Z_p$, where:\begin{equation}
p = \mathrm{GCD}(q, 5\omega), \hspace{4mm} p \neq 1
\end{equation}

The simplest case which satisfies these requirements is therefore the case where $p = 4$ and $5 \omega = 2$, corresponding to the Lens space $S_3/Z_4$ and an unbroken $Z_2$ symmetry once a non-zero $\mu$ term is generated. For our benchmark spectrum, we take these values and obtain from (\ref{gravitinoeq}), (\ref{VXeq}):\begin{equation}\label{gravitino2}
M_{3/2} \approx 35 \left(\frac{A_2}{Q}\right)\,\, \mathrm{TeV}.
\end{equation} Note that this also fixes $\eta$ in (\ref{GauginoMass.EQ}) \cite{Friedmann:2002ty}:\begin{equation}
\eta = 1 - \frac{5\alpha_{GUT}}{2 \pi} \ln\left(\frac{L(Q)}{q}\right) \approx 0.956
\end{equation}  We see that $1 - \eta$ is loop suppressed, and can only be $\mathcal{O}(1)$ for $q \gtrsim 100$. This validates our earlier claim that one naturally expects $\eta \sim 1$.

\begin{figure}
\centering
\includegraphics[scale=0.5]{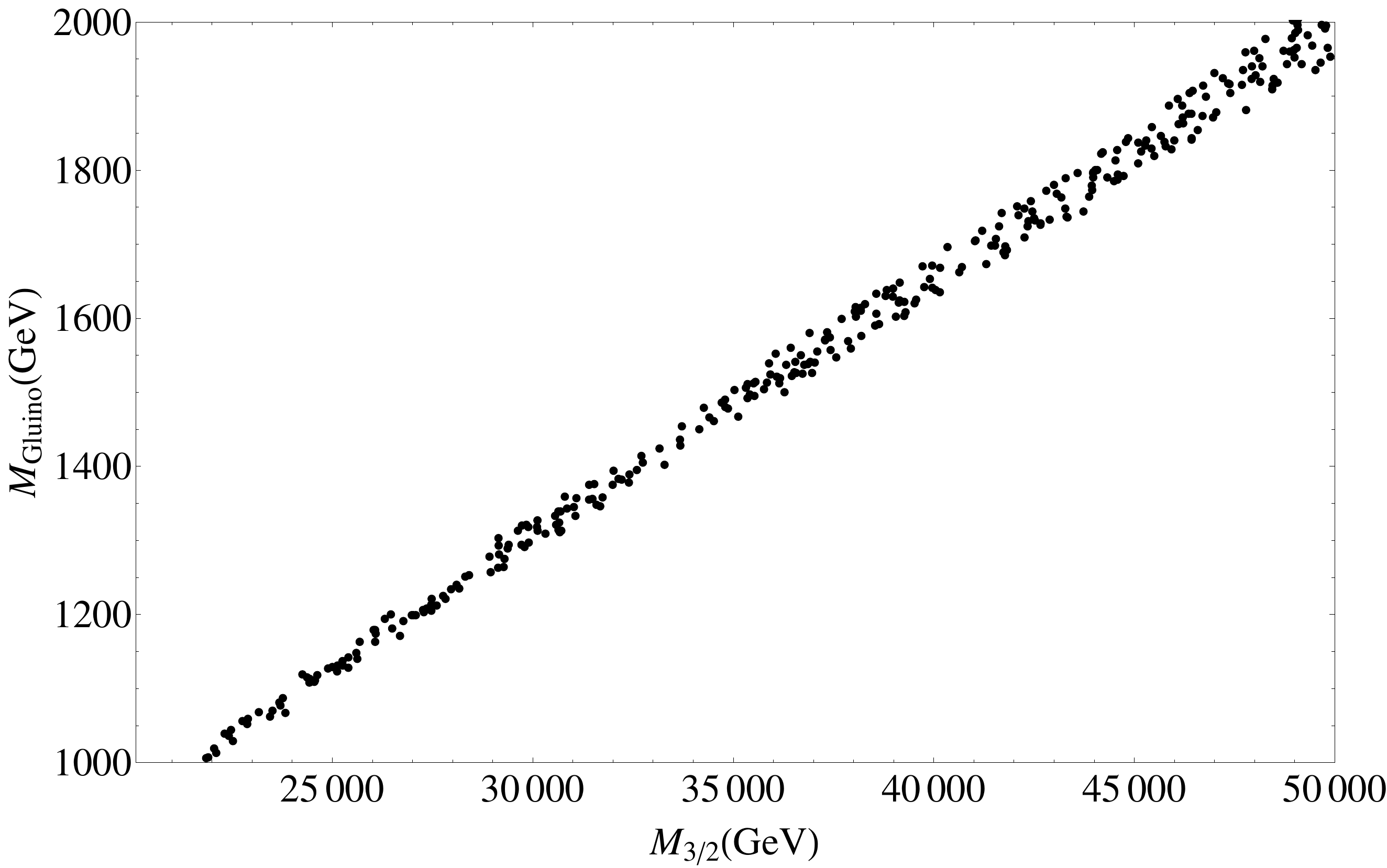}
\caption{Gluino mass vs $M_{3/2}$ for points which satisfy EWSB and Higgs mass constraints, i.e. the red points in Figure \ref{Fig1}.}
\label{Fig2}
\end{figure}

The only remaining undetermined factor in our benchmark spectrum is $A_2 / Q$ in (\ref{gravitino2}). In the context of Supersymmetric field theories, the precise normalization $A_2 = Q$ \cite{Affleck:1983mk} in the $\overline{DR}$ scheme was determined by requiring the consistency of various techniques used to study non-perturbative SUSY gauge theories \cite{Finnell:1995dr}. However, in the present context there may be UV threshold corrections which modify the relation $A_2/Q =1 $. To obtain the benchmark spectrum, we retain $A_2/ Q = 1$, which gives $M_{3/2} = 35$ TeV. Imposing constraints from EWSB and $M_h$ for this benchmark value fixes $C \approx 0.52$ and $\mu \approx 1.4$, resulting in the benchmark spectrum discussed in Section \ref{Spect.SEC}. Figure \ref{Fig2} illustrates the effect of relaxing this assumption; the gluino mass is plotted for $ 0.6 \lesssim A_2/Q \lesssim 1.4$, corresponding to $20\, \mathrm{TeV} \lesssim M_{3/2} \lesssim 50$ TeV. $M_{3/2} = 20$ TeV may barely avoid tension with BBN constraints \cite{Acharya:2010af}, depending on the particular values of moduli couplings and $T_{RH}$. 

In order to obtain a very approximate lower bound on $M_{\tilde{g}}$, we use the \verb FastLim  \cite{Papucci:2014rja} package. The program currently only implements some of the possible event topologies, and does not yet implement cascade decays such as $\tilde{g} \rightarrow q q \chi^0_2 \rightarrow q q h \chi^0_1$. Therefore, when using \verb FastLim  we treat the aforementioned cascade decays as $\tilde{g} \rightarrow q q \chi^0_2 \rightarrow q q \chi^0_1$, neglecting additional objects from $\chi^0_2 \rightarrow \chi^0_1 + h$. This simplification obviously reduces sensitivity to searches involving high ($\ge$ 6) jet multiplicity. Using this method, we obtain the approximate bound $M_{\tilde{g}} \gtrsim 1.1$ TeV in the compactified M-theory framework, though clearly a more precise analysis is desirable. Nonetheless, the benchmark spectrum with a 1.5 TeV gluino mass sits comfortably outside the excluded region, as we will discuss in Section \ref{Spect.SEC}.

\section{The spectrum and the branching ratios}
\label{Spect.SEC}

In this section we present the benchmark spectrum that results from the theoretical framework presented in Section \ref{Features.SEC}. In the compactified M-Theory all scalars are generically of order the gravitino mass, with the universal scalar mass given by $m_0 \sim \Order(1) M_{3/2}$ . We use the computing package \verb SOFTSUSY  \cite{Allanach:2001kg} to do two-loop RGE evolution of the high scale soft parameters to obtain sparticle pole masses.

\begin{figure}[h]
\centering
\begin{subfigure}{0.3 \textwidth}
\includegraphics[scale=0.31]{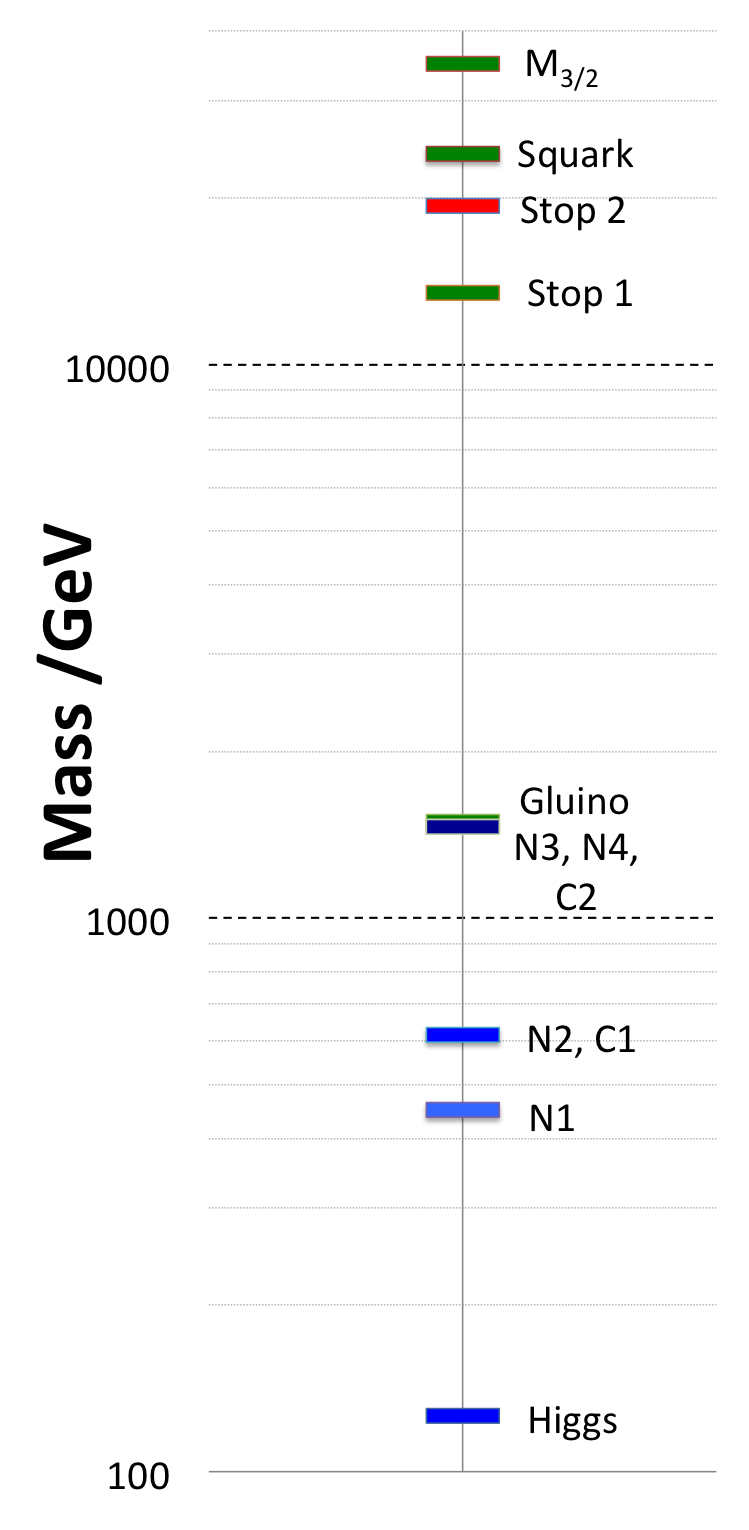}
\end{subfigure}
\hspace{2cm}
\centering
\begin{subfigure}{0.3 \textwidth}
\begin{tabular}{l || c}
Particle & Mass (GeV)\\
\hline
$m_0$ & 24200\\
$M_{3/2}$ & 35000 \\
$\sq_{L,R}$ & 24000 \\
$\tilde{t}_2$ & 19300 \\
$\tilde{t}_1$ & 13500 \\
$\tilde{b}_2$ & 23900 \\
$\tilde{b}_1$ & 19300 \\
$\sg$ & 1500 \\
$\chi_1^0$ & 450 \\
$\chi_2^0$ & 614 \\
$\chi_3^0$ & 1460 \\
$\chi_4^0$ & 1460 \\
$\chi_1^\pm$ & 614 \\
$\chi_2^\pm$ & 1460 \\
$h$  & 125.2\footnotemark \\
\end{tabular}
\end{subfigure}
\caption{Spectrum given GUT scale input values calculated from the theory for the central value $M_{3/2}=35 \TeV$. This spectrum has the GUT scale inputs $m_0 \approx 24$ TeV and $A_0 \approx 25$ TeV, where $m_0$ and $A_0$ are respectively the universal scalar mass and soft-breaking trilinear. The GUT scale gaugino masses are $M_1 = -1020 \GeV$, $M_2 = -730 \GeV$, $M_3=-590 \GeV$. Details on how this spectrum was derived are presented in Section \ref{Features.SEC}. For the gaugino masses and trilinear, we take the sign convention opposite to that of \texttt{SOFTSUSY}  \cite{Allanach:2001kg}. This relative sign affects the 2-loop term in the gaugino mass RGE's which is proportional to $A_t$.}
\label{Spect.FIG}
\end{figure}

We compute the branching ratios for the decays of some of the superpartners using \verb SDECAY  \cite{Muhlleitner:2003vg}, which are given in Table \ref{BRs.TAB}. We focus in particular on the superpartners that we expect to see at the LHC or at future colliders.
\begin{table}[h]
\tabcolsep=0.1cm 
\begin{minipage}{.1 \textwidth}
\centering
\begin{tabular}{l || c }
Decay & BR (\%)  \\
\hline
$\sg \to \chi_1^+ q_{1,2} \bar{q}_{1,2}$ & 25\\
$\sg \to \chi_1^\pm b\bar{t},~t\bar{b}$ & 23 \\
$\sg \to \chi_1^0 t \bar{t}$ & 20\\
$\sg \to \chi_2^0q_{1,2} {\bar{q}}_{1,2}$ & 12\\
$\sg \to \chi_1^0 q_{1,2} {\bar{q}}_{1,2}$ & 8 \\
$\sg \to \chi_2^0 b \bar{b}$& 7\\
$\sg \to \chi_2^0 t \bar{t}$ & 4\\
$\sg \to \chi_1^0 b \bar{b}$ & 1\\

\end{tabular}
\end{minipage}
\hspace{4cm}
\begin{minipage}{.1 \textwidth}
\begin{tabular}{l || c}
Decay & BR (\%)  \\
\hline
$\chi_4^0 \to \chi_1^\pm W^\mp$ & 60 \\
$\chi_4^0 \to \chi_2^0 h$ & 27 \\
$\chi_4^0 \to \chi_1^0 h$ & 8 \\
$\chi_4^0 \to \chi_2^0 Z$ & 4 \\
$\chi_4^0 \to \chi_1^0 Z$ & 2 \\
&\\
$\chi_3^0 \to \chi_1^\pm W^\mp$ & 60 \\
$\chi_3^0 \to \chi_2^0 Z$ & 26 \\
$\chi_3^0 \to \chi_1^0 Z$ & 8 \\
$\chi_3^0 \to \chi_2^0 h$ & 4\\
$\chi_3^0 \to \chi_1^0 h$ & 2 \\

& \\
$\chi_2^0 \to \chi_1^0 h$ & 98\\
$\chi_2^0 \to \chi_1^0 Z$ & 2\\
\end{tabular}
\end{minipage}
\hspace{4cm}
\begin{minipage}{.1 \textwidth}
\centering
\begin{tabular}{l || c}
Decay & BR (\%)  \\
\hline
$\chi_2^\pm \to \chi_1^\pm h$ & 31\\
$\chi_2^\pm \to \chi_1^\pm Z$ & 30\\
$\chi_2^\pm \to \chi_2^0 W^\pm$ & 30\\
$\chi_2^\pm \to \chi_1^0 W^\pm$ & 9\\
& \\
$\chi_1^\pm \to \chi_1^0 W^\pm$ & 100 \\

\end{tabular}
\end{minipage}
\caption{Branching ratios of gluino, neutralinos and charginos. The numbers don't add to 100 in the case of the $\chi_4^0$ branching ratios due to rounding errors.}
\label{BRs.TAB}
\end{table}
\footnotetext{For details regarding how the Higgs mass is calculated, we refer the reader to Appendix \ref{EWSB}.}

Since the squark masses are of $\Order(M_{3/2})$ at the high scale, they are not detectable at the LHC. RGE running splits the squarks to give the physical spectrum shown in Fig. \ref{Spect.FIG}. The neutralinos $\chi_1^0$ ($\chi_2^0$) are Bino (Wino)-like, while $\chi_3^0, \chi_4^0$ are Higgsino-like. Note that the mixing angles in the neutralino and chargino sectors are small, as $\mu \gg M_W, M_Z$. In previous studies of the $G_2$-MSSM, the LSP was taken to be Wino-like, as a Wino-LSP serves as a good DM candidate in string-motivated non-thermal cosmologies \cite{Moroi:1999zb}. However, obtaining a Wino-like as opposed to a Bino-like LSP requires large KK-threshold corrections to GUT scale gaugino masses (see e.g. \cite{Acharya:2012dz}), which we argued is unnatural in Section \ref{Features.SEC}. In this note, we assume that any would-be Bino-like LSP relic abundance decays to a hidden sector DM candidate, but the LSP is sufficiently long-lived to appear stable on collider scales\footnote{This occurs for example if $\chi^0_1$ decays to hidden sector DM via kinetic mixing through a $Z^\prime$, with $M_{Z^\prime} \sim \mathcal{O}(M_{3/2})$\cite{DMPaper}.}.

Of note is the 100 \% branching ratio of $\chi_1^\pm \to \chi_1^0 W^\pm$, and the 98.7 \% branching ratio of $\chi_2^0 \to \chi_1^0 h$. Note that the $\chi_2^0 \rightarrow \chi_1^0 Z$ decay width is subdominant to that of $\chi_2^0 \rightarrow \chi_1^0 h$. This can be explained as follows. The $\chi^0_2 \chi^0_1 h$ coupling arises from couplings of the form $\tilde{W}^0 \tilde{H} h$ and $\tilde{B}^0 \tilde{H} h$ in the gauge eigenstate basis. Since $\chi_1^0$ and $\chi_2^0$ are Bino and Wino-like, the $\chi_2^0 \rightarrow \chi_1^0 h $ amplitude is suppressed by a single power of gaugino-Higgsino mixing angles, $\mathcal{O}(M_Z/\mu)$. In contrast, in the gauge eigenstate basis only Higgsinos couple directly to $Z$ via couplings of the form $Z_\mu \tilde{H}^\dagger \overline{\sigma}^\mu \tilde{H}$. Thus the $Z_\mu {\chi_1^0}^\dagger \overline{\sigma}^\mu{\chi_2^0}$ coupling is suppressed by two powers of gaugino-Higgsino mixing angles, resulting in a suppresion of the $\chi_2^0 \rightarrow \chi_1^0 Z$ amplitude by $\mathcal{O}(M_Z^2/\mu^2)$. 

Note that this spectrum with a 1.5 TeV gluino should not have been discovered at LHC-8. In simplified models with decoupled squarks, the strongest limits on the gluino production come from multijet + MET searches \cite{Aad:2013a,Aad:2013wta,Chatrchyan:2014lfa,Chatrchyan:2013iqa}, which place the bound $M_{\tilde{g}} \gtrsim 1.35$ TeV. When setting limits on simplified models, the gluino is assumed to have a 100\% branching ratio into either $q q \chi_1^0$ or $t t \chi_10$ final states. For more realistic spectra such as the one considered here, the gluino mass bound will weaken significantly due to branching ratio factors.

\section{LHC-14 Predictions}
\label{LHC.SEC}

In this section, using the spectrum derived and presented above, we discuss the channels we expect to be observable at the LHC. We expect to see three channels, $pp \to \sg \sg$, $pp \to \chi_2^0 \chi_1^\pm$ and $pp \to \chi_1^\pm \chi_1^\mp$, and only these. The fact that only these three channels would be apparent is a feature of the M-theory construction. The scalars being heavy makes them kinematically inaccessible. The hierarchy between $\mu$ and the gaugino masses $M_a$ (and $\mu \gg M_Z$) results in a Bino-like LSP and Wino-like NLSP with heavy Higgsinos, meaning that only two neutralino/chargino direct production channels are accessible at LHC-14.

We computed using the production cross-sections to leading order for the three channels listed above using \verb MadGraph5  \cite{Alwall:2014hca} and multiplied them by K factors calculated using \verb Prospino  \cite{Beenakker:1996ed, Beenakker:1996ch, Beenakker:1999xh}. The results, including the expected number of events $N$ given 300 fb$^{-1}$ of data, are tabulated below. 

\begin{table}[H]
\centering
\begin{tabular}{l || c | c}
Channel & $\sigma$ (fb) & $N$ \\
\hline
$pp \to \sg \sg$ & 19 & 5800\\
$pp \to \chi_2^0 \chi_1^\pm$ & 19 & 5800\\
$pp \to \chi_1^\pm \chi_1^\mp$ & 10 & 3000 \\
\end{tabular}
\end{table}

Given the LSP mass of 450 GeV, a 1.5 TeV gluino is expected to be discoverable at the 5$\sigma$ level at LHC-14 given 300 fb$^{-1}$ of data \cite{CMS:2013xfa, Cohen:2013xda}. The gluino mass and cross-section allow immediate deduction of the gluino spin \cite{Kane:2008kw} and therefore confirmation that the discovery is indeed Supersymmetry. With squarks heavy, that procedure is straightforward. Furthermore, the $\chi^0_2 \rightarrow \chi^0_1 h $ decay mode allows for discovery potential in the chargino/neutralino direct production channels. For the benchmark spectrum considered here, direct $\chi^0_2, \chi^\pm_1$ production should be discoverable at LHC-14 with 1000 fb$^{-1}$ of data \cite{Baer:2012ts}. We understand that careful background studies need to be done to be sure these processes can be observed. The signatures are distinctive and event numbers large enough so it seems likely signals can be seen, but people more expert than us need to demonstrate the signals are really robust.

\section{Future Collider Predictions}
\label{FCC.SEC}

In this section, we briefly discuss the possible discoveries to be made at future colliders given the spectrum under consideration. We focus in particular on two possible proton-proton colliders, one with $\sqrt{s} = 50 \TeV$, and the other with $\sqrt{s} = 100 \TeV$. Prospects for Supersymmetry at such higher energy colliders has been studied recently \cite{Cohen:2013xda, Cohen:2014hxa, Low:2014cba}, although not in the context of a top-down, UV complete theory.

Given the spectrum, we find that some crucial new channels are accessible, namely $pp \to t \tilde{t}_1 \sg$, $pp \to b \tilde{b}_1 \sg$, $pp \to  \tilde{q}_{1 (L,R)} \sg$, $pp \to \chi_3^0 \chi_4^0$, $pp \to \chi_3^0 \chi_2^\pm$, $pp \to \chi_4^0 \chi_2^\pm$ and $pp \to \chi_2^\pm \chi_2^\mp$. 

Unfortunately, scalars are too heavy to be pair-produced \cite{Cohen:2014hxa}. However, we note that associated production of first family squarks with gluinos is accessible. The splitting of the light stop and the heavier first family squark masses, plus the kinematic measurement of the first family squarks, combine to give a precise measurement of the gravitino mass, the fundamental quantity that determines all masses in the theory, and thus deeply probes Supersymmetry breaking!

\begin{figure}[H]
\centering
\includegraphics[scale=0.3]{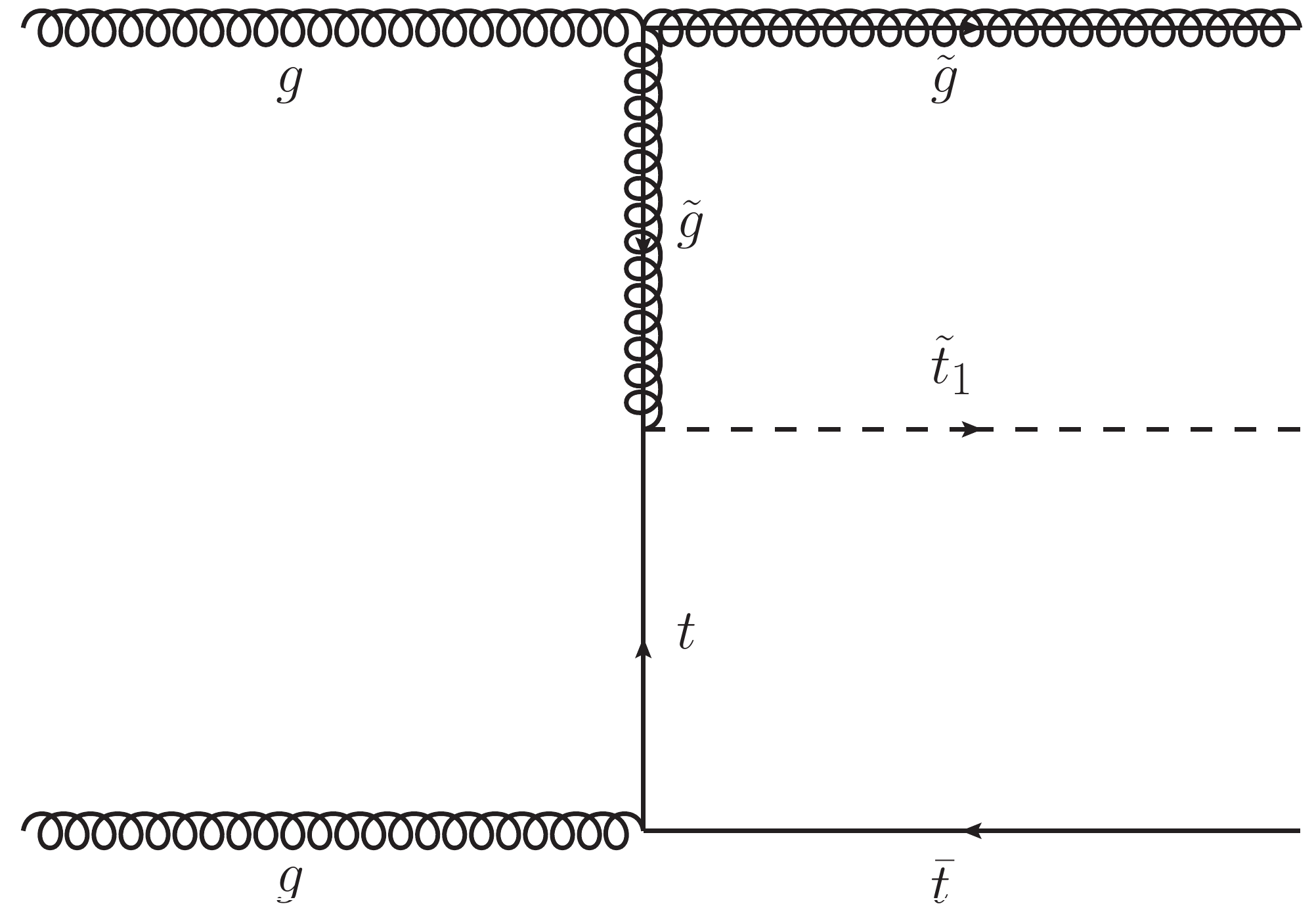}
\caption{Dominant Feynman graph for stop associated production by gluon splitting.}
\label{TStGo.FIG}
\end{figure}

 The associated production of stops and sbottoms can come from gluon splitting, or from the top (bottom) quark being considered as a parton. The question of when the top quark parton distribution function (PDF) becomes important has been studied recently \cite{Dawson:2014pea}, and it has been found that using a top PDF has only a small effect at $\sqrt{s} = 100 \TeV$. Therefore we simulate the production of stop gluino by looking at diagrams where the gluon splits and emits a top quark also. The bottom PDF is better known, and has a more significant effect at the energies in question, but for the purposes of this calculation, we present results using gluon splitting for the associated production of sbottoms. The dominant Feynman diagram for top stop gluino production is shown in fig. \ref{TStGo.FIG}. The diagram for sbottoms is identical with $b,\tilde{b}$ swapped for $t, \tilde{t}$. We recognise that careful studies of bottom and top PDFs need to be done to get fully reliable numbers.

We computed using \verb MadGraph5  \cite{Alwall:2014hca} the production cross-sections to leading order for these channels for both $\sqrt{s}=50 \TeV$ and 100 TeV. The results are tabulated below, including the number of events $N$ expected given 3000 fb$^{-1}$ of data.

\begin{table}[H]
\centering
\begin{tabular}{l || c | c | c | c}
Channel & $\sigma_{50~\TeV}$ (fb) & $N_{50~\TeV}$ & $\sigma_{100~\TeV}$ (fb) & $N_{100~\TeV}$ \\
\hline
$pp \to t \tilde{t}_1 \sg$ & $7.1\times 10^{-5}$ & 0 & $1.6\times 10^{-2}$ & 47\\
$pp \to b \tilde{b}_1 \sg$ & $2.6\times 10^{-6}$ & 0 & $3.0\times 10^{-3}$ & 9\\
$pp \to  \tilde{q}_{1(L,R)} \sg$ & $3.2\times 10^{-4}$ & 1 & $3.0 \times 10^{-1}$ & 900 \\
$pp \to \chi_3^0 \chi_4^0$ & $9.2\times 10^{-1}$ & 2800 & $3.4$ & 10200\\
$pp \to \chi_3^0 \chi_2^\pm$ & $1.8$& 5400 & $6.4$ & 19200\\
$pp \to \chi_4^0 \chi_2^\pm$ & $1.8$ & 5400 & $6.4$ & 19200\\
$pp \to \chi_2^\pm \chi_2^\mp$ & $1.0$ & 3000 & $3.7$& 11100\\
\end{tabular}
\end{table}
Although the number of stop events is not large, the gluino mass will already be known from LHC, which makes the stop and bottom search significantly simpler. The sbottom production cross-section is expected to be greater given that we make the approximation of gluon splitting rather than using the bottom PDF. We expect 15 or so stop events should already be found given 1000 fb$^{-1}$ of data, as well as 300 first generation squark events. The relatively large number of first family squark associated production events means these should be detectable. We also expect the heavy neutralinos and chargino to be detected, already at a 50 TeV, and certainly at a 100 TeV collider. Note there are other electroweakino production channels e.g. $\chi_2^0 \chi_2^\pm$ which have subdominant production cross-sections, two or more orders of magnitude smaller than those presented here. Because of their small production cross-sections, we do not list those channels here.

\section{Conclusion}
\label{Conclusion.SEC}

In this note, we have examined predictions for Supersymmetric particle masses in the $G_2$-MSSM, motivated by phenomenogically realistic compactifications of $M$-theory. By combining top-down constraints from moduli stabilization with bottom-up constraints from  EWSB and the measured Higgs mass, the sparticle spectrum is completely determined by $M_{3/2}$. Furthermore, given reasonable assumptions regarding the topology of the $G_2$ manifold, the gravitino mass is approximately calculable, giving a benchmark value $M_{3/2} = 35$ TeV.

We emphasize that we do not give a pure derivation of $M_{3/2} = 35$ TeV, or of an upper limit of $ \sim 50$ TeV. Rather we use all available top-down information from the UV theory, along with constraints and generic arguments, to find a ``natural" value of the gravitino and gluino masses in the $G_2$-MSSM. If one is agnostic about UV physics, one could by hand fine-tune these quantities away from their natural or generic values that we deduce from the $G_2$-MSSM plus constraints. The natural and generic values we report are quite different from the ``na\"ive natural" values in the absence of any theory.

The benchmark spectrum corresponding to $M_{3/2} = 35$ TeV is not constrained by LHC-8, and turns out to provide exciting phenomenology for LHC-14 and future colliders. The gluino mass is expected to be about 1.5 TeV for this benchmark spectrum, while the Wino(Bino)-like gaugino mass is about 614(450) GeV. The hierarchy between gaugino masses and $M_{3/2}$ arises because $M_{3/2}$ feels contributions from both the hidden sector meson and moduli F-terms, while gaugino masses only feel contributions from the moduli F-terms which are suppressed by about $\alpha_{GUT} \approx 1/25$ with respect to the meson F-terms. Three and only three production channels should discoverable at LHC-14: $pp \to \sg \sg$, $pp \to \chi_2^0 \chi_1^\pm$ and $pp \to \chi_1^\pm \chi_1^\pm$  where $\chi^0_1$ and $\chi^0_2$ are respectively Bino and Wino-like. The expected signature of the $\chi_1^\pm \chi_1^\pm$ channel is $\chi^+ \chi^- \rightarrow W^+ W^- + \, \mathrm{MET}$. The $\chi_2^0 \chi_1^\pm$ production channel gives the final state $\chi_2^0 \chi_1^\pm \rightarrow W^\pm\, h\, + \mathrm{MET}$, which should be quite a clear channel at the LHC \cite{Baer:2012ts}.

We have also investigated the prospects for the discovery of the heavier stops, first family squarks and Higgsinos at future colliders.  We find that associated production of gluino stop $p p \rightarrow \tilde{g} \tilde{t}_1 t$, gluino sbottom $p p \rightarrow \tilde{g} \tilde{b}_1 b$ as well as gluino squark production $p p \rightarrow \tilde{g} \tilde{u}, \tilde{d}$ should be seen at a 100 TeV collider, with leading-order production cross-sections of $1\times10^{-2}$, $3 \times 10^{-3}$ and $3\times10^{-1}$ fb respectively. This leads to hundreds of gluino-squark events given 3000 fb$^{-1}$ of data; precise knowledge of the gluino mass can help seperate these events from SM background. The heavy Higgsinos should also be detectable at a 50 TeV collider, and produced in relative abundance at a 100 TeV collider. The relevant Higgsino production channels are $pp \to \chi_3^0 \chi_4^0$, $pp \to \chi_3^0 \chi_2^\pm$,  $pp \to \chi_4^0 \chi_2^\pm$ and  $pp \to \chi_2^\pm \chi_2^\pm$. The relevant production cross sections at 50 (100) TeV are $\sigma \sim 1.8~ (6.4)$ fb for $pp \to \chi_{3,4}^0 \chi_2^\pm$, and $\sigma \sim 1.0~ (3.5)$ fb for $pp \to \chi_3^0 \chi_4^0$ and $pp \to \chi_2^\pm \chi_2^\pm$. Thus given 3000 fb$^{-1}$ of data, we expect of order a few thousand events for each channel at a 50 TeV collider, and of order tens of thousands of events at a 100 TeV collider.

To summarise, we have shown that the $G_2$-MSSM provides a constrained top-down framework, in which gluinos and some electroweakinos are discovered at LHC-14. The discovery of a single sparticle uniquely determines the remainder of the sparticle spectrum. Thus given a discovery of gauginos at LHC-14, a discovery of  squarks and Higgsino at 100 TeV colliders within the predicted mass range would give strong evidence towards Supersymmetry and an UV completion like the compactified M-theory construction presented here.

\section*{Acknowledgements}
We would like to thank B.S. Acharya, J.R. Ellis, P. Kumar, M. Papucci, A. Pierce and C.P. Yuan for useful discussions. The work of SARE, GLK and BZ is supported in part by Department of Energy grant DE-SC0007859.

\begin{appendix}

\section{Appendix A: Moduli Stabilization and SUSY Breaking in the $G_2$-MSSM}\label{moduli}

In this Appendix, we review the relationship between low energy Supersymmetry and UV dynamics within the $G_2$-MSSM. This relationship is discussed in \cite{Acharya:2008zi,Acharya:2008hi}, and was preceded by a detailed study of dynamical Supersymmetry breaking through moduli stabilization \cite{Acharya:2006ia,Acharya:2007rc}. We will only describe results relevant for our goal of making predictions for Supersymmetric spectra for LHC and future colliders. Interested readers are referred to \cite{Acharya:2012tw} for a more detailed summary. 

In order to stabilize moduli and obtain a de Sitter vacuum, previous studies have assumed the generic presence of an $SU(Q) \times SU(P+1)$ hidden sector, where $SU(Q)$ is pure super Yang-Mills and $SU(P+1)$ is super QCD-like with $N_f = 1$. Non-perturbative effects generate the following superpotential \cite{Affleck:1983mk,Affleck:1984xz} which breaks Supersymmetry and stabilizes all moduli:
\begin{equation}\label{super}
W = A_1 \phi^{-2/P} \exp{ i b_1 f_1} + A_2 \exp{i b_2 f_2}
\end{equation} where $2 \pi b_1\,(2 \pi b_2) = P^{-1}  (Q^{-1})$. Here $f_1$ and $f_2$ are the gauge kinetic functions of the $SU(Q)$ and $SU(P+1)$ gauge groups, and $\phi$ is the $SU(P+1)$ meson condensate in Planck units. We assume $f_1 \propto f_2$; otherwise it is difficult to stabilize moduli within the supergravity approximation \cite{Acharya:2007rc}. For consistency with previous works, we make the simplifying assumption $f_1 = f_2 = f_{vis}$, where $f_{vis}$ is the visible sector gauge kinetic function.  We furthermore assume that the visible sector is an $SU(5)$ GUT.

In order to proceed with the moduli stabilization analysis, we use the K\"ahler potential derived in \cite{Acharya:2008hi}:\begin{equation}\label{Kahler}
K = - 3 \log 4 \pi^{1/3} V_X + \kappa_{\alpha\beta}\frac{\Phi_\alpha^\dagger \Phi_\beta}{V_X} + \frac{\overline{\phi} \phi}{V_X} + c_{\alpha \beta} \frac{\overline{\phi} \phi}{V_X} \frac{\Phi^\dagger_\alpha \Phi_\beta}{V_X} + ...
\end{equation} 
where the volume of the $G_2$ manifold, $V_X$, is a homogenous function of the moduli of degree 7/3 \cite{Beasley:2002db}, $\Phi$, $\Phi^\dagger$ represent visible sector fields, $\kappa_{\alpha\beta}$ is the K\"ahler metric, and ``..." represent higher dimensional operators which are neglected in the analysis. 

The term proportional to $c_{\alpha \beta}$ is a higher order correction studied in \cite{Acharya:2008hi}. It induces flavor changing neutral currents unless the elements of $c_{\alpha \beta}$ are all small, or $c_{\alpha \beta} \sim\propto \kappa_{\alpha\beta}$ up to small corrections. Because scalars are heavy and the tree-level CP phases are rotated away, flavor effects are not expected to cause phenomenogical problems \cite{Kadota:2011cr}. For simplicity, we assume here that the coefficients of $c_{\alpha \beta}$ are diagonal and universal in the basis where $\kappa_{\alpha \beta}$ is diagonal. Thus we take:\begin{equation}\label{higherorder}
c_{\alpha \beta} = \left(\frac{C}{3}\right) \kappa_{\alpha \beta}
\end{equation} The normalization for $C$ in (\ref{higherorder}) follows that of \cite{Acharya:2008hi}. We will see that $C \neq 0$ is required for consistent EWSB. Relaxing the assumption taken in (\ref{higherorder}) will induce moderate flavor dependence in the soft breaking parameters, but will not change the gravitino mass or the overall sparticle mass scale.

Given (\ref{super}),(\ref{Kahler}) and (\ref{higherorder}), determining the dynamics of moduli stabilization is straightforward, though rather cumbersome and has been carried out in \cite{Acharya:2008hi}. The gravitino mass is determined by the standard supergravity relation:\begin{equation}\label{m32}
M_{3/2} = m_{pl} e^{K/2} \left| W\right| \approx 0.035\, {V_X}^{-3/2} m_{pl} \left|Q - P \right| \frac{A_2}{Q} e^{-\left(\frac{P_{eff}}{Q-P}\right)} , \hspace{3mm} P_{eff} = \frac{ 14 (3 (Q-P ) - 2)}{3 ( 3 ( Q - P) - 2 \sqrt{6 (Q - P)})}
\end{equation} where in the second equality we have fixed moduli at their stabilized values and imposed vanishing vacuum energy \cite{Acharya:2007rc,Acharya:2008hi}. Note that for typical values, $V_X \sim \mathcal{O}(730)$ \cite{Acharya:2008hi}. Imposing vanishing of the vacuum energy enforces $Q - P \ge 3$, but does not fix $Q - P$ itself. However, from (\ref{m32}) we see that for $Q - P \ge 4$, there still exists a large hierarchy between $ M_{3/2}$ and the electroweak scale. Thus if a vanishing vacuum energy is imposed, the set of solutions with $Q - P = 3$ solves the hierarchy problem. Taking $Q - P = 3 $ in (\ref{m32}) results in:\begin{equation}
M_{3/2} \approx \frac{9 \times 10^{5}}{V_X^{3/2}} \left(\frac{A_2}{Q}\right)\, \mathrm{TeV}.
\end{equation} This result was used in Section \ref{benchmark} to obtain the benchmark value $M_{3/2} \approx 35$ TeV.

Using the standard relations between the Supergravity Lagrangian and soft SUSY breaking parameters \cite{Brignole:1997dp}, it is then straightforward to determine the relation between soft scalar masses, trilinears and $M_{3/2}$, as well as the gaugino masses $M_a$ ($a=1,2,3$) at the renormalization scale $Q = M_{GUT}$:\begin{equation}\label{softparams}
m_0^2 \approx M_{3/2}^2 \left( 1 - C\right),\hspace{2mm} A^0 \approx 1.5 M_{3/2}\left(1 - C\right), \hspace{2mm} M_a = \frac{e^{K/2} F^i \partial_i f_{vis}}{2 i\, \mathrm{Im} f_{vis}} + M_a^{anomaly} \end{equation} where $m_0$ is the universal scalar soft mass, and we haved define trilinears as $\mathcal{L} = - y_{ij} A\, \tilde{Q}_{i,L} H_q \tilde{q}_{j,R}$. We have seperated the tree-level and anomaly-induced \cite{Bagger:1999rd} contributions to $M_a$. The tree level contribution to $M_a$ is proportional to the moduli F-terms $F_s$; moduli stabilization results in $F_s \propto M_{3/2}/V_Q$ where $V_Q \approx 1/\alpha_{GUT}$ is the visible sector 3-cycle volume \cite{Acharya:2008zi}. Thus, the tree-level gaugino masses are $M^{tree}_a \sim \alpha_{GUT}~ M_{3/2}$, and since the anomaly contribution is loop-suppressed, the two contributions are of comparable size. Such a hierarchy between scalars and gaugino masses has also been observed in other corners of string theory \cite{Conlon:2006us, Choi:2007ka}, though the physical mechanism may be different.

Including both $M_a^{tree}$ and $M_a^{anomaly}$, the full expression for the gaugino masses upon moduli stabilization is given by \cite{Acharya:2008hi}:
\beq
M_a \approx \left[-0.032 \eta  +\alpha_{GUT}\left(0.034 \left(3 C_a - C^\prime_a \right) + 0.079 C^\prime_a(1 - C)\right) \right]\times M_{3/2}
\label{GauginoMass.EQ}
\eeq
where $C_a = (0, 2, 3)$ and $C_a^\prime = (33/5, 7, 6)$ for $U(1)_Y$, $SU(2)_L$ and $SU(3)_c$ respectively. Here $1 - \eta$ parameterizes threshold corrections to the $SU(5)$ gauge coupling from Kaluza-Klein modes; as discussed in Section \ref{benchmark}, $\left|1 - \eta\right| \lesssim 0.1$ is expected for generic compactifications.

Thus taking $\eta \approx 1$, we have shown that top-down constraints from moduli stabilization fix the sparticle spectrum in terms of three almost-calculable constrained quantities: $M_{3/2}, \, \mu, \,$ and $C$. Note that all SUSY breaking parameters in (\ref{softparams}), (\ref{GauginoMass.EQ}) are implicitly defined at the renormalisation scale $Q= M_{GUT}$. Given these ``high scale" values, one must take into account RG-evolution and other radiative corrections to obtain sparticle pole masses as discussed in Section \ref{Spect.SEC}.

\section{Appendix B: Constraints from EWSB and $M_h$}\label{EWSB}

In this appendix, we discuss how both EWSB and with consistency with $M_h = 125.2 \pm 0.4$ GeV are imposed to reduce the dimensionality of the $M_{3/2}, \mu, C$ parameter space. We begin by consider the EWSB conditions, given by:\begin{align}\label{EWSB1}
& \frac{M_Z^2}{2} = \frac{\overline{m}_{H_d}^2 - \overline{m}_{H_u}^2 \tan^2\beta}{\tan^2\beta - 1} - \mu^2 \\ \label{EWSB2}
& B\mu = \frac{1}{2} \sin 2 \beta \left(\overline{m}_{H_u}^2 + \overline{m}_{H_d}^2 + 2 \mu^2 \right)
\end{align} where $\overline{m}_{H_u}$ and $\overline{m}_{H_d}$ are the tadpole corrected $H_u,\, H_d$ soft masses, and all parameters are evaluated at the renormalization scale $Q_{EWSB}^2 = m_{\tilde{t}_1} m_{\tilde{t}_2}$. For the M-theory models we consider, we expect $\mu \lesssim 0.1 M_{3/2}$. 

There are two independent arguments for such a suppression of $\mu$ with respect to $M_{3/2}$. The first is a top-down argument, which is related to the doublet-triplet splitting mechanism \cite{Witten:2001bf} discussed in Section \ref{benchmark}. This mechanism results in a geometric symmetry which forbids the $\mu$ term, which is broken by moduli stabilization to generate $\mu$ via the Giudice-Masiero mechanism \cite{Giudice:1988yz}. Thus generically $\mu$ is suppressed with respect to $M_{3/2}$ by moduli \emph{vev}'s \cite{Acharya:2011te}. This suppression was estimated by \cite{Acharya:2011te} to be roughly an order of magnitude, though current theoretical uncertainties preclude a precise estimate. Another argument for $\mu \lesssim 0.1 \,M_{3/2}$ is motivated by electroweak naturalness. A measure of fine-tuning in EWSB is the degree to which the two terms on the right hand side of (\ref{EWSB1}) are required to cancel in order to obtained the measured value of $M_Z$. Thus electroweak naturalness favors $\mu^2 \ll {M_{3/2}^2}$, which mitigates the cancellation in (\ref{EWSB1}) required to obtain $M_Z \approx 90$ GeV \cite{Feldman:2011ud}.

In addition to $\mu \lesssim 0.1 \,M_{3/2}$, we also expect $0 < \overline{m}_{H_u}^2 \ll \overline{m}_{H_d}^2$, as $m_{H_u}^2$ runs significantly between $Q = M_{GUT}$ to $Q = Q_{EWSB}$ due to the top Yukawa coupling, while the running of $m_{H_d}^2$ is small for moderate $\tan \beta \lesssim 10$. The lower bound $0 < \overline{m}_{H_u}^2$ is required for consistency with $\mu \lesssim 0.1 M_{3/2}$, as can be seen from (\ref{EWSB1}). Taking the $B \mu, \overline{m}_{H_d}^2 \gg \mu^2, \, \overline{m}_{H_u}^2$ and $\sin 2 \beta \approx 2/\tan \beta$ limit, (\ref{EWSB1}) can be written as \cite{Feldman:2011ud}:\begin{equation}
\label{EWSBapprox} \mu^2 \approx \frac{\overline{m}_{H_d}^2}{B^2 - \overline{m}_{H_d}^2}\left(\overline{m}_{H_u}^2 + \frac{M_Z^2}{2}\right)\end{equation} where $B \approx 2 \, M_{3/2}$ at $Q = M_{GUT}$ \cite{Acharya:2008zi} and $ B \approx 1.7 M_{3/2}$ at $Q = Q_{EWSB}$ \cite{Feldman:2011ud}. Taking $\overline{m}_{H_d}^2 \approx m_0^2$ and $\overline{m}^2_{H_u} = {M^2_{3/2}} f(C)$ where $f(C)$ accounts for the running of $m_{H_u}$ due to the top trilinear $A_t$, we can recast (\ref{EWSBapprox}) in a more suggestive form:\begin{equation}\label{EWSBconstraint}
\mu^2 \approx \frac{{M_{3/2}^2} (1-C)}{B^2 - {M_{3/2}^2} (1-C)} \left( {M_{3/2}^2}\,f(C) + \frac{M_Z^2}{2}\right).\end{equation} Note that $f(C)$ decreases monotonically as $C$ increases \cite{Feldman:2011ud}. Thus we have used EWSB conditions to obtain a constraint on the allowed $M_{3/2}, \mu, C$ space; we use one-loop RGE's and one-loop effective potential corrections \cite{Pierce:1996zz} to compute $f(C)$ for Figure \ref{Fig1}. In order to minimize large logarithmic corrections, we run $\alpha_{s}$ and $y_t$ using a 2-step procedure which explicity accounts for the decoupling of squarks at $Q = Q_{EWSB}$.

We now discuss how the constraint on $M_h$ is incoporated in our analysis. The mass of the lightest Higgs boson in the MSSM decoupling limit is given schematically by:\begin{equation}\label{Mh}
M_{h}^2 = M_Z^2 \cos^2 2 \beta + \delta M_h^2
\end{equation} where $\delta M_h^2$ denotes radiative corrections from both SM and MSSM particles; the value of $\delta M_h^2$ is fixed for given values of $M_{3/2}, \mu, C$. We can then use the EWSB condition (\ref{EWSB2}) in the  $B \mu, \overline{m}_{H_d}^2 \gg \mu^2, \, \overline{m}_{H_u}^2$ and $\sin 2 \beta \approx 2/\tan \beta$ limit to express $\tan \beta$ in terms $M_{3/2}, C$ and $\mu$:\begin{equation}\label{mutanbeta}
\tan \beta \approx \frac{m_0^2}{B \mu} \approx \frac{M_{3/2} (1 - C)}{B \mu}
\end{equation} 

Thus combining (\ref{Mh}) and (\ref{mutanbeta}) with $M_h = 125.2 \pm 0.4$ GeV, we obtain an additional constraint in the $M_{3/2}, \mu, C$ parameter space. To compute $M_h$ for given $M_{3/2}, \mu, C$, we use the ``match-and-run" procedure, outlined for example in \cite{Giudice:2011cg,Kane:2011kj}. Note that some of the authors here made an error in \cite{Kane:2011kj} regarding SM radiative corrections to $M_h$; correcting this error increases $M_h$ by $\sim 1.5$ GeV. Our calculation here uses 3-loop RGE's and 2-loop threshold corrections for the matching procedure \cite{Draper:2013oza}. From Figure \ref{Fig1}, we see that the slice of parameter space consistent with $M_h = 125.2 \pm 0.4$ has a non-negligible width. This is due predominantly to experimental uncertanties in $M_t, \alpha_s$ and $M_h$ \cite{Kane:2011kj}; more precise measurements of these quantities will sharpen constraints on this parameter space.
\end{appendix}


\begin{thebibliography}{15}
\bibitem{CMS:2014ega} 
  CMS Collaboration [CMS Collaboration],
  ``Precise determination of the mass of the Higgs boson
and studies of the compatibility of its couplings with the standard model,''
  CMS-PAS-HIG-14-009.
\end{thebibliography}
\end{document}